\documentclass[11pt,preprint]{aastex}
\usepackage{rotating}
\shortauthors{Stolte et al.}
\shorttitle{Innershell Photoionization}

\begin{document}
\doublespace

\title{Innershell Photoionization Studies of Neutral Atomic Nitrogen}

\author{W. C. Stolte\altaffilmark{1,2},
V. Jonauskas\altaffilmark{3},
D. W. Lindle\altaffilmark{1},
M. M. Sant'Anna\altaffilmark{4},
D. W. Savin\altaffilmark{5}}
\altaffiltext{1}{Department of Chemistry, University of Nevada, Las
  Vegas, NV 89154, USA}
\altaffiltext{2}{Advanced Light Source, Lawrence Berkeley National
  Laboratory, Berkeley, CA 94720, USA}
\altaffiltext{3}{Institute of Theoretical Physics and Astronomy,
  Vilnius University, Go\v{s}tauto St. 12, LT-01108 Vilnius,
  Lithuania}
\altaffiltext{4}{Instituto de F\'{\i}sica, Universidade Federal do Rio
  de Janeiro, 21941-972 Rio de Janeiro, Brazil}
\altaffiltext{5}{Columbia Astrophysics Laboratory, Columbia
  University, New York, NY 10027, USA}

\begin{abstract}

Innershell ionization of a $1s$ electron by either photons or
electrons is important for X-ray photoionized objects such as active
galactic nuclei and electron-ionized sources such as supernova
remnants.  Modeling and interpreting observations of such objects
requires accurate predictions for the charge state distribution (CSD)
which results as the $1s$-hole system stabilizes.  Due to the
complexity of the complete stabilization process, few modern
calculations exist and the community currently relies on 40-year-old
atomic data.  Here, we present a combined experimental and theoretical
study for innershell photoionization of neutral atomic nitrogen for
photon energies of $403-475$~eV.  Results are reported for the total
ion yield cross section, for the branching ratios for formation of
N$^+$, N$^{2+}$, and N$^{3+}$, and for the average charge state.  We
find significant differences when comparing to the data currently
available to the astrophysics community.  For example, while the
branching ratio to N$^{2+}$ is somewhat reduced, that for N$^+$ is
greatly increased, and that to N$^{3+}$, which was predicted not to be
zero, grows to $\approx 10\%$ at the higher photon energies studied.
This work demonstrates some of the shortcomings in the theoretical CSD
data base for innershell ionization and points the way for the
improvements needed to more reliably model the role of innershell
ionization of cosmic plasmas.

\end{abstract}


\section{Introduction}
\label{sec:Intro}

Innershell ionization of a $1s$ electron by either photons or
electrons can be an important process for a range of cosmic sources.
These include X-ray photoionized objects such as active galactic
nuclei or X-ray binaries \citep{Kall01a}, photoionization of
interstellar dust grains \citep{Dwek96a}, and shock-driven gas such as
in supernova remnants \citep{Vink12a}.  The resulting $1s$-hole system
relaxes via a complicated cascade of fluorescence and/or Auger-ejected
electrons.  This process affects both the emitted spectrum of a source
as well as the ionization structure of the gas.

Driven by the astrophysical importance of innershell ionization,
\citet{Kaas93a} carried out the first comprehensive study into the
resulting fluorescence and Auger yields for all ionization stages of
the elements from beryllium to zinc.  These data are derived from
theoretical calculations published between 1969-1972 using
$LS$-coupling and a central field approximation.  The data are also
presented using configuration-averaging, thereby providing no fine
structure information for the fluorescence or Auger processes.

Given the understandable limitations of the \citet{Kaas93a} data, in
the past decade, there has been significant theoretical effort using
relativistic $LSJ$ methods to calculate the fluorescence yield due to
an initial $1s$ hole
\citep[e.g.,][]{Baut03a,Gorc03a,Palm03a,Palm03b,Mend04a,Garc05a,
  Gorc06a,Haso06a,Palm08a,Palm08b,Garc09a,Palm11a,Palm12a}.  That work
finds a number of differences with \citet{Kaas93a}.  For example,
fluorescence channels which were thought to be closed are in fact open
and channels which are open can have fluorescence yields which differ
by factors of several or more.

There is far less theoretical work on the resulting charge state
distribution.  This is not surprising given the complexity of the
complete stabilization process.  Innershell ionization of iron, for
example, can result in the ejection of up to 8 additional electrons
\citep{Kall01a}.  However, this multiple Auger process is often
ignored by astrophysicists, probably because the needed
state-of-the-art data are lacking.  This omission likely hinders our
astrophysical understanding of cosmic sources as multiple electron
emission can dramatically change the predicted ionization structure of
an astrophysical source, particularly in ionizing plasmas
\citep{Hahn15a}.

To begin to address this matter, here we report a combined theoretical
and experimental study into innershell photoionization on neutral
atomic nitrogen.  The rest of this paper is organized as follows.  In
Section~\ref{sec:Theoretical} we briefly review the theoretical method
used.  The experimental method is presented in
Section~\ref{sec:Experimental}.  The theoretical and experimental
results are presented and discussed in Section~\ref{sec:Results} .
Lastly, we give a summary of our findings in
Section~\ref{sec:Summary}.

\section{Theoretical Approach}
\label{sec:Theoretical}

Our photoionization calculations account for both the promotion of a
$1s$ electron into the continuum and also resonant photoexcitation of
a $1s$ electron.  The former leads directly to ionization, but both
processes contribute to the total ionization cross section through a
sequential cascade of radiative and Auger transitions as the system
relaxes.  In addition we account for direct ionization of an electron
from higher shells.

We have calculated energy levels, radiative and Auger transition
probabilities, and photoionization cross sections using the Flexible
Atomic Code (FAC), which employs the Dirac-Fock-Slater method
\citep{Gu08a}.  The total number of levels included is 3781.  We used
1787 levels for N, 1470 for N$^{+}$, 480 for N$^{2+}$, and 44 for
N$^{3+}$.  The necessary bound and continuum wavefunctions for each
ion were calculated using the local central potential for the ground
configuration.  The wavefunctions are obtained separately for each
charge state using self-consistent field procedures.  As a result, the
wavefunctions for different charge states are not orthogonal.  The
calculated photoionization cross sections and Auger transition rates
involve using wavefunctions from different charge states.  The
non-orthogonality of the wavefunctions for the different ionization
states is expected to have only a small influence on these
calculation.  When calculating the relevant matrix elements, the
additional terms introduced are small, often have different signs, and
approximately cancel out when summed \citep{Cowa81a}.  This is
sometimes referred to as ``compensation effects''.  The mixing of all
configurations belonging to the same ionization stage is taken into
account.

Level populations have been estimated in every step of the cascade
process using \citep{Pala10a,Jona11a}
\begin{equation}
\label{eq:nj}
N_{j}=\sum_{i} N_{i} \frac{A_{ij}}{A_i},
\end{equation}
where $N_{j}$ is the population of level $j$, $N_i$ is the population
of the initial level $i$, $A_{ij}$ is the Auger or radiative
transition probability from $i$ to $j$, and $A_{i}$ is the total decay
rate of $i$, from both Auger and radiative transitions.  The initial
level populations due to photoionization and photoexcitation at a
given photon energy have been obtained from the corresponding cross
sections divided by the total cross section summing all processes at
that energy.

At first we performed cascade calculations including all Auger and
radiative transitions.  Those calculations were time consuming and
generated very large data files but demonstrated that the radiative
cascade contribution to the ion yield is negligible.  The fluorescence
yield is less than 4\% for all configurations with a K-shell vacancy
except for two levels of the $1s 2s 2p^3$ with yields of about 9\%.
These small fluorescence yields result in autoionization branching
ratios close to unity.  Hence, we have neglected radiative cascades in
our subsequent calculations.  This resulted in a reduced computation
time and yielded smaller data files.

We also accounted for additional ionization due to the sudden
change of the ion charge after photoionization, a process known as
shake-off.  For this we used the sudden perturbation approximation,
which is expected to be valid for high energy photons when the
resulting initial photoelectron leaves the system at high velocity and
does not have time to interact with the remaining bound electrons.
For an initial vacancy with principle quantum number $n_0$ and
angular momentum $l_0$, the probability $P$ for ionization of an
electron from the $nl^{w}$ shell, where $w$ is the number of electrons
in that shell, is given by \citep{Jona09a}
\begin{equation}
P(K^\prime \rightarrow K^{\prime\prime}) =
w (1-\langle nl_{K} | nl_{K^\prime} \rangle^2)
\langle nl_{K} | nl_{K^\prime} \rangle^{2(w-1)}.
\label{shake}
\end{equation}
$K$ is the initial configuration of the system and the electrons from
the $nl$ shell are described by the radial wavefunction of $| nl_{K}
\rangle$.  $K^\prime$ is the subsequent configuration produced by the
vacancy in the $n_{0}l_{0}$ shell and $K^{\prime\prime}$ results when
an electron is removed from the $nl$ shell of the $K^\prime$
configuration.  For our work here $n_0l_0 = 1s$.
Equation~(\ref{shake}) describes the rearrangement of the complex as
it transitions from the $K$ to $K^\prime$ system and additional
electrons are shaken off.  Here the shake-off probability is
determined using the configuration-averaged approximation.

In order to obtain the level populations for the $K^{\prime\prime}$
configuration, we calculated the population transfer due to shake-off
transitions using
\begin{equation}
\label{eq:shake2}
N_{K^{\prime\prime}_{j}} =
P(K^\prime \rightarrow K^{\prime\prime})
\left[\frac{2 J_{j}+1}{g(K'')}\right]
H(\varepsilon_{ph} - [E_{K^{\prime\prime}_{j}} - E_{\rm g}])
\sum_{K^\prime_{i}} N_{K^\prime_{i}}.
\end{equation}
Here $N_{K^{\prime\prime}_{j}}$($N_{K^{\prime}_{i}}$) is population of
level $j$($i$) for the $K^{\prime\prime}$($K^{\prime}$) configuration,
$J_{j}$ is the total angular momentum quantum number of level $j$,
$g(K^{\prime\prime})$ is the statistical weight of configuration
$K^{\prime\prime}$, $E_{\rm g}$ is the energy of the ground level of
neutral atom, and $H(x)$ is the Heaviside step function which ensures
that only energetically possible shake-off transitions are taken into
account.  The summation is performed over all levels of the $K^\prime$
configuration.

Our calculated shake-off probability for the N$^+(1s^1
2s^2 2p^3)$ configuration, due to a forming a $1s$ vacancy in the
initial N$(1s^{2} 2s^{2} 2p^{3})$ atom, is 17\% for transition to a
N$^{2+}(1s^{1} 2s^{2} 2p^{2})$ configuration and 6\% for transition to
N$^{2+}(1s^{1} 2s^{1} 2p^{3})$ configuration.  Either of these two
configurations can further decay through Auger transitions to form
N$^{3+}$.  This is the only pathway to forming N$^{3+}$ through
photoionization of N as the direct triple photoionization is 
expected to be unimportant.

\section{Experimental Method}
\label{sec:Experimental}

The measurements were performed on the undulator beamline 8.0.1.3 at
the Advanced Light Source located at Lawrence Berkeley National
Laboratory, in Berkeley, CA, USA.  The basic experimental setup has
been previously described in \citet{Stol08a}, as has the method used
to generate atomic species \citep{Sant11a}, which is similar to that
used for the creation of atomic chlorine \citep{Stol13a} and atomic
oxygen \citep{Stol97a,McL13a}.

In short, atomic N was generated by flowing commercially obtained
99.9995\% pure gaseous N$_2$ through a microwave discharge cavity.
The products of this nitrogen plasma were a mixture of atoms and
molecules in ground and excited states. A combination of a fast gas
flow rate with teflon and phosphorus pentoxide coatings on the flow
tubes, in addition to tubing shape and discharge distance, reduced
wall recombination effects and strongly quenched nitrogen atoms
created in the long-lived metastable states ($1s^22s^22p^3\ [^2{\rm
    D}^o,\ ^2{\rm P}^o]$).  A final enhancement was the addition of a
constant magnetic field, satisfying the electron-cyclotron resonance
condition, being perpendicularly superimposed on the the 2.45 GHz
electric field of the microwave cavity.  The resulting mixture of
$1s^22s^22p^3\ [^4S^o]$ ground-state nitrogen atoms and molecular
N$_2$ were constrained to flow through a small orifice of 0.5~mm
diameter before entering the interaction region to interact with the
X-rays.

Partial ion yields were measured using the method described in the
references above and includes a mixture of contributions from both
ground-state atomic N and molecular N$_2$.  Although we
eventually subtract it out, the presence of a molecular signal within
our spectra provides an excellent internal energy calibration
\citep{Chen89a,Seme06a}. The cross section, $\sigma^{q+}(E)$, as a
function of photon energy, $E$, for photoionization of atomic nitrogen
to an ion of charge $+q$ can be obtained from \citep{Sams85a}
\begin{equation}
\sigma^{q+}(E) = C_{q^+} (I_{\rm on}^{q+} - fI_{\rm off}^{q+}),
\label{eq:ion}
\end{equation}
where $I_{\rm on}^{q+}$ and $I_{\rm off}^{q+}$ are normalized ion
yields measured as a function of photon energy with the microwave
discharge on or off, respectively.  $C_{q+}$ is a constant dependent
on the number density of nitrogen atoms and the ion-collection
efficiency of the apparatus.  Absolute data for single and multiple
photoionization of N$_2$ from \citet{Stol98a} were used to determine
values for the constants C$_{q+}$.  The parameter $f = \rho
{(\rm{N}_2^{\rm on})/} \rho (\rm{N}_2^{\rm off})$ represents the
fraction of N$_2$ molecules which do not dissociate in the discharge,
with $\rho(\rm{N}_2^{\rm on})$ and $\rho(\rm{N}_2^{\rm off})$ being
number densities of N$_2$ with the microwave discharge on or off,
respectively.  The value of $f$ is empirically chosen to eliminate the
molecular peaks from the measured ion yields via a weighted
subtraction \citep{Sams86a,Sams90a,Stol97a}.  As noted above, the
dissociation fraction is $1-f$, or approximately 4\% here.  Finally,
the collection efficiency for each ion N$^{q+}$ produced by
photoionization of atomic nitrogen was assumed to be equal to the
collection efficiency of the same N$^{q+}$ generated by dissociative
photoionization of N$_2$.

\section{Results and Discussion}
\label{sec:Results}

\begin{figure}
\center \includegraphics[width=0.6\textwidth]{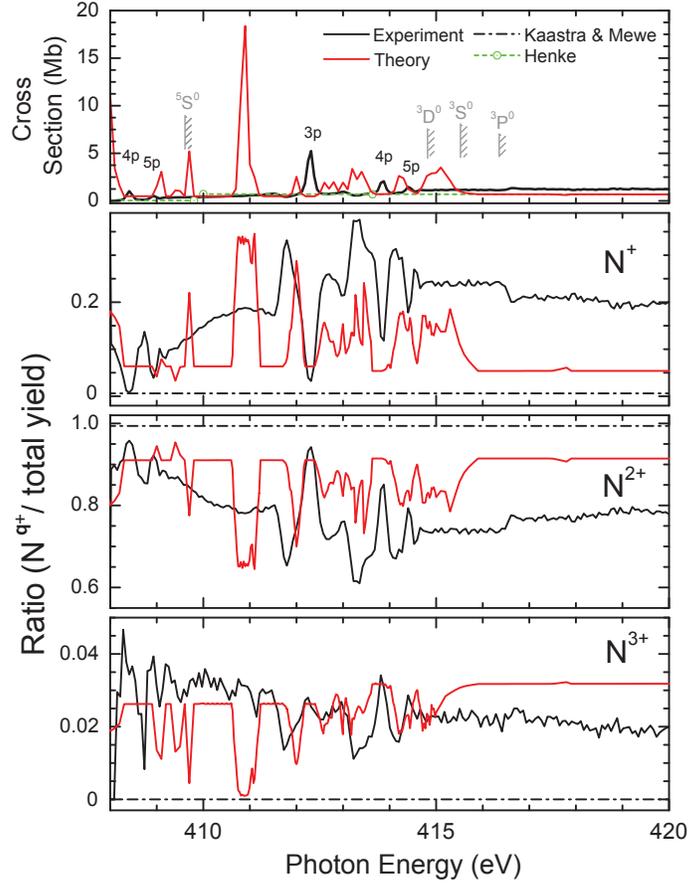}
\caption{\label{fig:Ratio2} The top panel shows the photoionization
  cross section for total ion yield (TIY) of atomic N near the
  $1s$-ionization threshold.  Various core excitations and associated
  autoionizing resonances are labeled.  Our experimental results are
  shown in black (with an estimated uncertainty of $12\%$ in magnitude
  and $\sim 50$~meV in energy).  Our theoretical results are in red.
  Also shown are the measurements of \cite{Henke1993} in green.  The
  corresponding branching ratios for forming N$^{q+}$ relative to the
  TIY are shown in the lower three panels.  The published theoretical
  results of \citet{Kaas93a} are shown by the dot-dashed lines.}
\end{figure}

\begin{figure}
\center \includegraphics[width=1\textwidth]{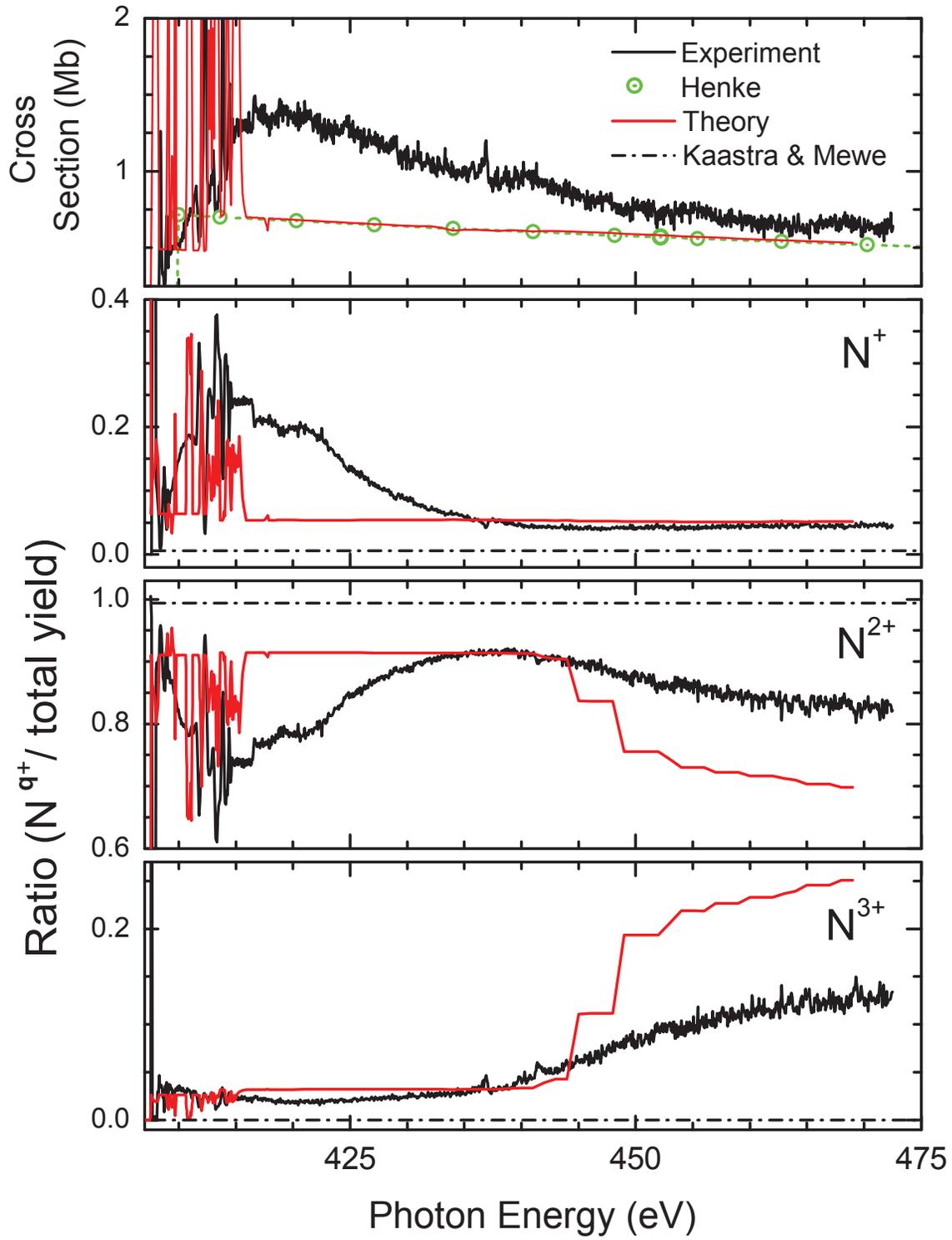}
\caption{\label{fig:Ratio1} Same as Figure~\ref{fig:Ratio2} but from
  $410-475$~eV.}
\end{figure}

\begin{figure}
\center
\includegraphics[width=1\textwidth]{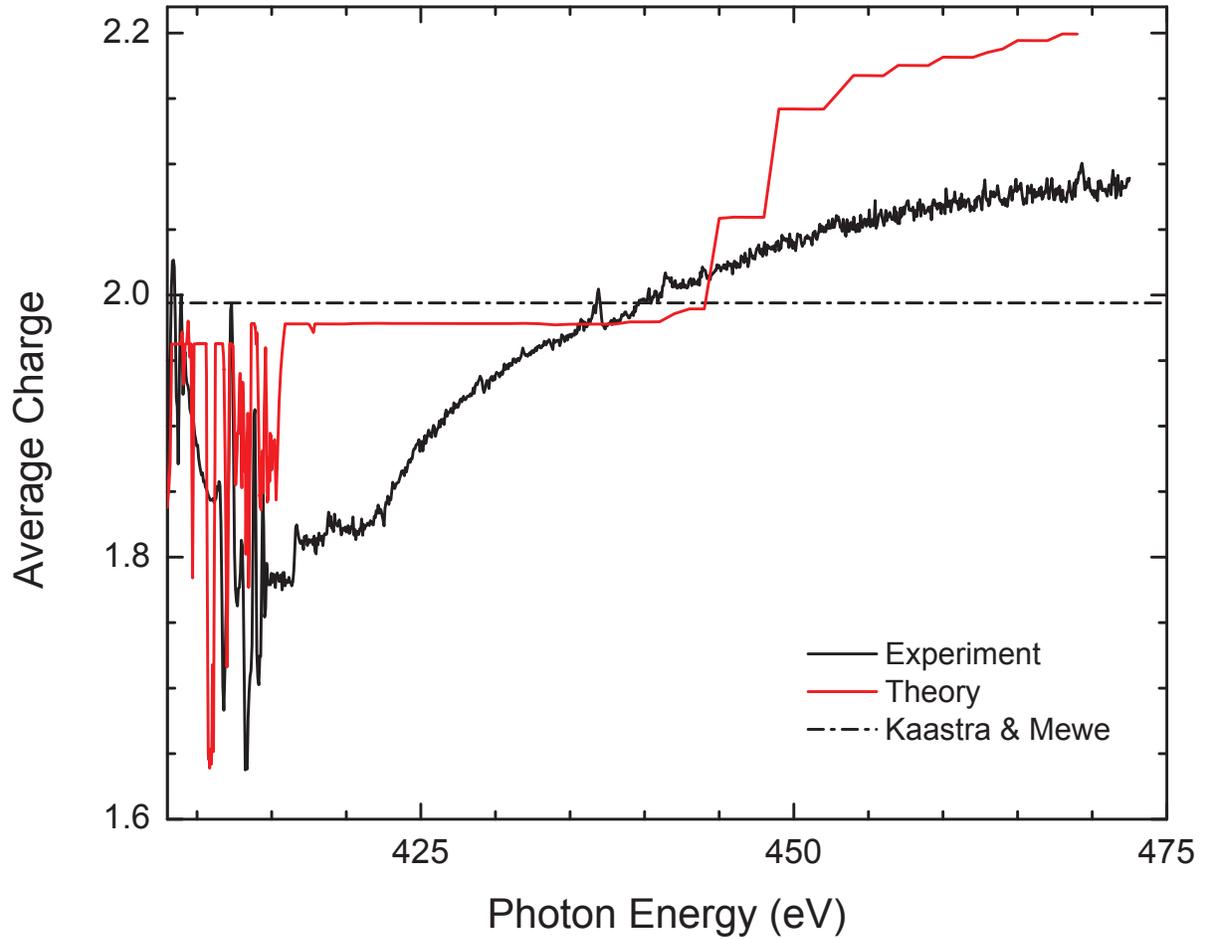}
\caption{\label{fig:AverageCharge} Average nitrogen charge state
  following photoionization of atomic nitrogen.  The black curve
  presents the experimental data, the red curve represents our present
  theoretical results, and the dot-dashed line shows the published
  theoretical results of \citet{Kaas93a}.}
\end{figure}

In Figure~\ref{fig:Ratio2} we present our measured and theoretical
cross sections for the total ionization yield (TIY) between
$408-420$~eV, which is near the the $1s$-ionization threshold.  Using
our experimental and theoretical results, we have computed the
branching ratio for forming N$^{q+}$ relative to the total atomic ion
yield.  Figure~\ref{fig:Ratio1} shows our results over the larger
energy range of $410-475$~eV.  Also shown in these figures are the
experimental cross section results of \citet{Henke1993} and the
theoretical branching ratio results of \citet{Kaas93a}.  In
Table~\ref{tab:henke} we present our experimental and theoretical
total photoionization cross sections along with the experimental
results of \citet{Henke1993}.  Table \ref{tab:ratios} presents the
branching ratio data at selected photon energies, along with the
average final charge state, which is also shown in
Figure~\ref{fig:AverageCharge}.

Comparing our experimental TIY cross section data to that of
\citet{Henke1993}, we find a significant difference between the two.
This should not be surprising since the previous results were
calculated from molecular nitrogen data taken at fixed photon energies
using lamp sources and then dividing those absolute cross sections by
two.  This estimate of dividing the molecular value by two should
become valid with high energy, but as can be seen in
Figure~\ref{fig:Ratio1}, it is still nearly 20$\%$ lower 55 eV above
threshold.  We also note that only one actual measured point at
452.2~eV from \citet{Henke1993} lies within our energy region.  The
rest are calculated results using dispersion equations, empirical
relations to directly relate the incident photon energy with atomic
photoabsorption cross section.  As a result of the coarse energy grid
used by \citet{Henke1993}, none of the resonance structure was
observed at the time.

We find a rich resonance structure in our data.  In the $408-415$~eV
range, the measured cross section shows several clear resonances due
to core-excited autoionizing $1s2s^22p^3np$ resonances ($n=3-5$)
leading up to the various series limits for the core excitation
\citep{Sant11a}.  These show up in the branching ratios as dips in the
N$^+$ branching ratio, whereas they tend to appear as peaks in the
N$^{2+}$ and N$^{3+}$ ratios.  Interestingly, the branching ratios
show an even greater variation than is seen in the cross section.
This rapidly varying behavior continues until the photon energy
converges to the ${\rm ^3D^o}$, ${\rm ^3S^o}$, and ${\rm ^3P^o}$
series limits for the core excitation, above which the behavior is
smoother.  We note also that the resonances above the $1s$-ionization
threshold, the ${\rm ^5S^o}$ limit at 409.64 eV, show clear
Fano-Beutler profiles.  This indicates that there are strong
interferences between the core-excited intermediate resonance states
and the open continua.  These profiles show up as valleys in the N$^+$
ratio and as peaks in the N$^{2+}$ and N$^{3+}$ ratios.

Our calculated first ionization threshold energy is 407.49~eV, which
is 2.15~eV smaller than the experimental value.  Similarly, the
measured resonance energies lie about 2~eV below our theoretical
values.  We attribute these differences to missing correlation effects
in the calculations.  This is not surprising as the configuration
interaction method, which is widely used by the community, converges
very slowly for neutral atoms, making it very difficult for theory to
obtain good agreement with experiment.  Another noticeable difference
is the strength of the $1s 2s^{2} 2p^{3} 3p$ resonance which is higher
in our calculations and the energy which is lower.  It is interesting
that current calculations for cross sections are in the better
agreement with the measurements of \citet{Henke1993}.  We attribute
this to the lack of correlation effects in both works.  

Overall, better agreement between theory and experiment for the TIY
cross section was found for the R-matrix calculations of
\citet{Sant11a}.  However, the approach used there was unable to
generate branching ratios, which are important for modeling cosmic
plasmas.

Of particular interest is the baseline behavior in the N$^+$ and
N$^{2+}$ branching ratios from threshold to about 435~eV.  Over this
range the experimental N$^+$ branching ratio starts above theory but
then decreases into good agreement while the N$^{2+}$ measurements
start below theory but then increases into good agreement.  This
behavior is not predicted by our theory and the cause for the
discrepancy is unclear.  It is unlikely to be due to post-collision
interactions (PCIs) which can occur just above a photoionization
threshold.  Such interactions could prevent a slow photoelectron from
actually escaping from the system, thus reducing the N$^{2+}$ yield
and enhancing the N$^+$ yield for the first few eV above an ionization
threshold.  However, PCI is predicted not to be strong enough to
account for the magnitude of the effect seen.  Any variation in the
partial cross section for L-shell photoionization is also unlikely to
be the cause.  We are over an order of magnitude in energy above the
L-shell ionization threshold and the L-shell partial cross section is
predicted to be smooth in this energy range.  It seems to us that the
more likely cause is variation in the partial cross section for
K-shell photoionization due to states which have not been included in
the calculations.

Moving on to N$^{3+}$, the branching ratio yield for this ion starts
out very low just above the K-edge.  At these energies N$^{3+}$ forms
via the creation of a $1s$ hole followed by double-Auger decay.  Other
processes are even less likely, and also beyond our computational
capabilities.  The N$^{3+}$ branching ratio then begins to increase
around 435 eV due primarily to direct double-ionization (${\rm K +
  L}$) followed by a single-Auger decay.  The first resonance in this
Rydberg series is located at $436.9 \pm 0.1$~eV and is predicted by
our calculations to be due to the $1s 2s^2 2p^2 3p$ configuration.
The series limit is near 444~eV.  This resonance can be seen in the
total cross section, in the N$^{3+}$ branching ratio, to a lesser
lesser extent in the N$^{2+}$ data, and by a dip in the N$^{+}$
results.  Unfortunately, due to the small cross section and the
measurement noise levels, we are unable to perform a Rydberg analysis
to experimentally determine these ionization thresholds.

For N$^{3+}$, we also find a discrepancy between our experimental and
theoretical results.  The measured N$^{3+}$ branching ratio increases
slower than predicted and correspondingly the N$^{2+}$ results
decrease slower than predicted.  We attribute this to the theory which
is expected to overestimate double photoionization cross sections at
these energies.  The shake-off approach utilized here to describe the
double photoionization process is valid for sudden changes in the
potential of the system.  The observed differences between experiment
and theory demonstrates that the change of the system charge does not
occur quite as quickly at these photon energies as the approach
requires.

Looking at all three final charge states produced, we see that
eventually the branching ratios and final average charge state all
appear to reach asymptotic values at about 470~eV.  The only processes
which are possible at higher energies, are double-K-shell ionization
and direct triple-ionization (${\rm K+L+L}$), both of which are
expected to be negligibly small.

Comparing to the results of \cite{Kaas93a} we find significant
differences in the branching ratios and average final charge state.
Given the theoretical calculations that their work is derived from, it
is not surprising that they do not include any of the observed
resonance structure and energy dependence.  These differences remain
even where the experimental results appear to have reached their
asymptotic limit.  Our measured branching ratios are larger than
\citet{Kaas93a} for forming N$^+$ and smaller for forming N$^{2+}$.
Moreover, we find a significant branching ratio for forming N$^{3+}$,
a channel which was considered to be closed by \citet{Kaas93a}.  Thus
it appears that K-shell ionization processes can increase the average
charge state of a cosmic plasma faster than is currently predicted by
models.  Exploring the full astrophysical implications of our
experimental findings, which will require implementing our data into
various astrophysical models, is beyond the scope of our work here.

\section{Summary}
\label{sec:Summary}

We have reported innershell photoionization measurements and
calculations of atomic nitrogen near the $1s$-ionization threshold.
Significant differences are found with the recommended data currently
used by the astrophysics community, particularly for the final charge
state distribution.  For example, the branching ratio for N$^+$
formation is significant greater than given in the data base and as
well as that for N$^{3+}$, a channel which the recommended data
predict is closed.  These results point out some of the shortcomings
in the 40-year-old theoretical data currently available for the CSD
due to innershell ionization.  Generating the state-of-the-art data
necessary in order to more reliable modeling of cosmic sources will
require a concerted theoretical and experimental effort covering all
systems where a $1s$-hole can be formed.

\acknowledgements

The authors thank J.~Kaastra, T.~R.~Kallman, R.~K.~Smith, and
E.~M.~Gullikson for stimulating conversations.  WCS and DWL wish to
acknowledge support by the National Science Foundation under NSF Grant
No.\ PHY-09-70125.  MMS was supported by the CNPq-Brazil.  DWS was
supported in part by the NASA Astrophysics Research and Analysis
Program.  The experimental portion of this work was performed at the
Advanced Light Source, which is supported by DOE (DE-AC03-76SF00098).

\newpage

\bibliography{ArticleFile}

\begin{thebibliography}{35}
\expandafter\ifx\csname natexlab\endcsname\relax\def\natexlab#1{#1}\fi

\bibitem[{Bautista {et~al.}(2003)Bautista, Mendoza, Kallman, \&
  Palmeri}]{Baut03a}
Bautista, M.~A., Mendoza, C., Kallman, T.~R., \& Palmeri, P. 2003, Astron. \&
  Astrophys., 403, 339

\bibitem[{Chen {et~al.}(1989)Chen, Ma, \& Sette}]{Chen89a}
Chen, C.~T., Ma, Y., \& Sette, F. 1989, Phys. Rev. A, 40, 6737

\bibitem[{Cowan(1981)}]{Cowa81a}
Cowan, R.~D. 1981, {The Theory of Atomic Structure and Spectra} (Berkeley:
  University of California Press, 1981)

\bibitem[{Dwek \& Smith(1996)}]{Dwek96a}
Dwek, E. \& Smith, R.~K. 1996, Astrophys. J., 459, 686

\bibitem[{Garc\'{\i}a {et~al.}(2009)Garc\'{\i}a, Kallman, Witthoeft, Behar,
  Mendoza, Palmeri, Quinet, Bautista, \& Klapisch}]{Garc09a}
Garc\'{\i}a, J., Kallman, T.~R., Witthoeft, M., Behar, E., Mendoza, C.,
  Palmeri, P., Quinet, P., Bautista, M.~A., \& Klapisch, M. 2009, Astrophys. J.
  Suppl. Ser., 185, 477

\bibitem[{Garc\'{\i}a {et~al.}(2005)Garc\'{\i}a, Mendoza, Bautista, Gorczyca,
  Kallman, \& Palmeri}]{Garc05a}
Garc\'{\i}a, J., Mendoza, C., Bautista, M.~A., Gorczyca, T.~W., Kallman, T.~R.,
  \& Palmeri, P. 2005, Astrophys. J. Suppl. Ser., 158, 68

\bibitem[{Gorczyca {et~al.}(2006)Gorczyca, Dumitriu, Haso{\u{g}}lu, Korista,
  Badnell, Savin, \& Manson}]{Gorc06a}
Gorczyca, T.~W., Dumitriu, I., Haso{\u{g}}lu, M.~F., Korista, K.~T., Badnell,
  N.~R., Savin, D.~W., \& Manson, S.~T. 2006, Astrophys. J. Lett., 638, L121

\bibitem[{Gorczyca {et~al.}(2003)Gorczyca, Kodituwakku, Korista, Zatsarinny,
  Badnell, Behar, Chen, \& Savin}]{Gorc03a}
Gorczyca, T.~W., Kodituwakku, C.~N., Korista, K.~T., Zatsarinny, O., Badnell,
  N.~R., Behar, E., Chen, M.~H., \& Savin, D.~W. 2003, Astrophys. J., 592, 636

\bibitem[{Gu(2008)}]{Gu08a}
Gu, M.~F. 2008, Can. J. Phys., 86, 675

\bibitem[{Hahn \& Savin(2015)}]{Hahn15a}
Hahn, M. \& Savin, D.~W. 2015, Astrophys. J.

\bibitem[{Haso{\u{g}}lu {et~al.}(2006)Haso{\u{g}}lu, Gorczyca, Korista, Manson,
  Badnell, \& Savin}]{Haso06a}
Haso{\u{g}}lu, M.~F., Gorczyca, T.~W., Korista, K.~T., Manson, S.~T., Badnell,
  N.~R., \& Savin, D.~W. 2006, Astrophys. J. Lett., 649, L149

\bibitem[{Henke {et~al.}(1993)Henke, Gullikson, \& Davis}]{Henke1993}
Henke, B.~L., Gullikson, E.~M., \& Davis, J.~C. 1993, At. Data Nucl. Data
  Tables, 54, 181

\bibitem[{Jonauskas {et~al.}(2009)Jonauskas, Ku\v{c}as, \& Karazija}]{Jona09a}
Jonauskas, V., Ku\v{c}as, S., \& Karazija, R. 2009, Lith. J. Phys. Tech., 49,
  415

\bibitem[{Jonauskas {et~al.}(2011)Jonauskas, Ku\v{c}as, \& Karazija}]{Jona11a}
---. 2011, Phys. Rev. A, 84, 53415

\bibitem[{Kaastra \& Mewe(1993)}]{Kaas93a}
Kaastra, J.~S. \& Mewe, R. 1993, Astron. \& Astrophys. Suppl. Ser., 97, 443

\bibitem[{Kallman \& Bautista(2001)}]{Kall01a}
Kallman, T. \& Bautista, M. 2001, Astrophys. J. Suppl. Ser., 133, 221

\bibitem[{McLaughlin {et~al.}(2013)McLaughlin, Ballance, Bowen, Gardenghi, \&
  Stolte}]{McL13a}
McLaughlin, B.~M., Ballance, C.~P., Bowen, K.~P., Gardenghi, D.~J., \& Stolte,
  W.~C. 2013, Astrophys. J. Lett., 771, L8

\bibitem[{Mendoza {et~al.}(2004)Mendoza, Kallman, Bautista, \&
  Palmeri}]{Mend04a}
Mendoza, C., Kallman, T.~R., Bautista, M.~A., \& Palmeri, P. 2004, Astron. \&
  Astrophys., 414, 377

\bibitem[{Palaudoux {et~al.}(2010)Palaudoux, Lablanquie, Andric, Ito,
  Shigemasa, Eland, Jonauskas, Ku\v{c}as, Karazija, \& Penent}]{Pala10a}
Palaudoux, J., Lablanquie, P., Andric, L., Ito, K., Shigemasa, E., Eland,
  J.~H.~D., Jonauskas, V., Ku\v{c}as, S., Karazija, R., \& Penent, F. 2010,
  Phys. Rev. A, 82, 43419

\bibitem[{Palmeri {et~al.}(2003{\natexlab{a}})Palmeri, Mendoza, Kallman, \&
  Bautista}]{Palm03a}
Palmeri, P., Mendoza, C., Kallman, T.~R., \& Bautista, M.~A.
  2003{\natexlab{a}}, Astron. \& Astrophys., 403, 1175

\bibitem[{Palmeri {et~al.}(2003{\natexlab{b}})Palmeri, Mendoza, Kallman,
  Bautista, \& Mel\'{e}ndez}]{Palm03b}
Palmeri, P., Mendoza, C., Kallman, T.~R., Bautista, M.~A., \& Mel\'{e}ndez, M.
  2003{\natexlab{b}}, Astron. \& Astrophys., 410, 359

\bibitem[{Palmeri {et~al.}(2008{\natexlab{a}})Palmeri, Quinet, Mendoza,
  Bautista, Garc\'{\i}a, \& Kallman}]{Palm08a}
Palmeri, P., Quinet, P., Mendoza, C., Bautista, M.~A., Garc\'{\i}a, J., \&
  Kallman, T.~R. 2008{\natexlab{a}}, Astrophys. J. Suppl. Ser., 177, 408

\bibitem[{Palmeri {et~al.}(2008{\natexlab{b}})Palmeri, Quinet, Mendoza,
  Bautista, Garc\'{\i}a, Witthoeft, \& Kallman}]{Palm08b}
Palmeri, P., Quinet, P., Mendoza, C., Bautista, M.~A., Garc\'{\i}a, J.,
  Witthoeft, M.~C., \& Kallman, T.~R. 2008{\natexlab{b}}, Astrophys. J. Suppl.
  Ser., 179, 542

\bibitem[{Palmeri {et~al.}(2011)Palmeri, Quinet, Mendoza, Bautista,
  Garc\'{\i}a, Witthoeft, \& Kallman}]{Palm11a}
---. 2011, Astron. \& Astrophys., 525, A59

\bibitem[{Palmeri {et~al.}(2012)Palmeri, Quinet, Mendoza, Bautista,
  Garc\'{\i}a, Witthoeft, \& Kallman}]{Palm12a}
---. 2012, Astron. \& Astrophys., 543, A44

\bibitem[{Samson \& Angel(1990)}]{Sams90a}
Samson, J.~A.~R. \& Angel, G.~C. 1990, Phys. Rev. A, 42, 1307

\bibitem[{Samson \& Pareek(1985)}]{Sams85a}
Samson, J.~A.~R. \& Pareek, P.~N. 1985, Phys. Rev. A, 31, 1470

\bibitem[{Samson {et~al.}(1986)Samson, Shefer, \& Angel}]{Sams86a}
Samson, J.~A.~R., Shefer, Y., \& Angel, G.~C. 1986, Phys. Rev. Lett., 56, 2020

\bibitem[{Sant'Anna {et~al.}(2011)Sant'Anna, Schlachter, \"{O}hrwall, Stolte,
  Lindle, \& McLaughlin}]{Sant11a}
Sant'Anna, M.~M., Schlachter, A.~S., \"{O}hrwall, G., Stolte, W.~C., Lindle,
  D.~W., \& McLaughlin, B.~M. 2011, Phys. Rev. Lett., 107, 33001

\bibitem[{Semenov {et~al.}(2006)Semenov, Cherepkov, Matsumoto, Fujiwara, Ueda,
  Kukk, Tahara, Sunami, Yoshida, Tanaka, Nakagawa, Kitajima, Tanaka, \&
  DeFanis}]{Seme06a}
Semenov, S.~K., Cherepkov, N.~A., Matsumoto, M., Fujiwara, K., Ueda, K., Kukk,
  E., Tahara, F., Sunami, T., Yoshida, H., Tanaka, T., Nakagawa, K., Kitajima,
  M., Tanaka, H., \& DeFanis, A. 2006, J. Phys. B, 39, 375

\bibitem[{Stolte {et~al.}(2013)Stolte, {Felfli Z. Guillemin}, \"{O}hrwall, Yu,
  Young, Lindle, Gorczyca, Deb, Manson, Hibbert, \& Msezane}]{Stol13a}
Stolte, W.~C., {Felfli Z. Guillemin}, R., \"{O}hrwall, G., Yu, S.-W., Young,
  J.~A., Lindle, D.~W., Gorczyca, T.~W., Deb, N.~C., Manson, S.~T., Hibbert,
  A., \& Msezane, A.~Z. 2013, Phys. Rev. A, 88, 53425

\bibitem[{Stolte {et~al.}(2008)Stolte, Guillemin, Yu, \& Lindle}]{Stol08a}
Stolte, W.~C., Guillemin, R., Yu, S.-W., \& Lindle, D.~W. 2008, J. Phys. B, 41,
  145102

\bibitem[{Stolte {et~al.}(1998)Stolte, He, Cutler, Lu, \& Samson}]{Stol98a}
Stolte, W.~C., He, Z.~X., Cutler, J.~N., Lu, Y., \& Samson, J.~A.~R. 1998, At.
  Data Nucl. Data Tables, 69, 171

\bibitem[{Stolte {et~al.}(1997)Stolte, Lu, Samson, Hemmers, Hansen, Whitfield,
  Wang, Glans, \& Lindle}]{Stol97a}
Stolte, W.~C., Lu, Y., Samson, J.~A.~R., Hemmers, O., Hansen, D.~L., Whitfield,
  S.~B., Wang, H., Glans, P., \& Lindle, D.~W. 1997, J. Phys. B, 30, 4489

\bibitem[{Vink(2012)}]{Vink12a}
Vink, J. 2012, Annu. Rev. Astron. Astrophys., 20, 49

\end{thebibliography}

\newpage

\begin{table}
\centering
\begin{tabular}{|c|c|c|c|} 
\hline
Photon Energy (eV) & \multicolumn{3}{|c|}{Cross Section (Mb)} \\
\cline{2-4} 
                   & Experiment & Theory & Henke et al. \\
\hline
409.80 & 0.539 & 0.487 & 0.027 \\
410.00 & 0.531 & 0.487 & 0.714 \\
413.64 & 0.905 & 0.816 & 0.700 \\
420.33 & 1.35  & 0.682 & 0.674 \\
427.12 & 1.20  & 0.654 & 0.650 \\
434.03 & 1.00  & 0.612 & 0.626 \\
441.05 & 0.955 & 0.607 & 0.603 \\
448.19 & 0.775 & 0.595 & 0.581 \\
452.20 & 0.712 & 0.581 & 0.569 \\
455.44 & 0.748 & 0.571 & 0.560 \\
462.80 & 0.648 & 0.549 & 0.540 \\
470.29 & 0.637 & 0.531 & 0.520 \\
\hline
\end{tabular}
\\
\caption{Present experimental and theoretical total photoionization
  cross section of atomic N along with the published results of
  \citet{Henke1993}.  The cross sections are given in units of
  Megabarns (Mb) which correspond to $10^{-18}$~cm$^2$.  Our estimated
  combined statistical and systematic uncertainty on the experimental
  results is 12$\%$ for the cross section and $\approx 50$~meV for the
  photon energy.}
\label{tab:henke}
\end{table}

\newpage

\begin{sidewaystable} [h]
\centering
\begin{tabular}{|c|*{12}{c|}} \hline
Photon & \multicolumn{9}{|c|}{Branching Ratio ($\%$)} & \multicolumn{3}{|c|}{Average Final Charge State} \\ \cline{2-13}
Energy & \multicolumn {3}{|c|} {Experiment} & \multicolumn {3}{|c|} {Theory} & \multicolumn{3}{|c|}{Kaastra \& Mewe} & Experiment & Theory & Kaastra \& Mewe \\ \cline{2-10}
(eV) & N$^+$ & N$^{2+}$ & N$^{3+}$ & N$^+$ & N$^{2+}$ & N$^{3+}$ & N$^+$ & N$^{2+}$ & N$^{3+}$ & & & \\
\hline

$ \ge 403$   &      &      &      &      &      &      & 0.6 & 99.4 & 0.0 & & & 1.99 \\
403   & 94.2 & 5.5* & 0.3* & 92.3 & 7.7  & 0     & & & & 1.06 & 1.08 & \\
417   & 21.0 & 76.9 & 2.1  & 5.36 & 91.4 & 3.18 & & & & 1.81 & 1.98 & \\
420   & 20.1 & 77.9 & 2.0  & 5.36 & 91.4 & 3.18 & & & & 1.82 & 1.98 & \\
426   & 12.8 & 84.8 & 2.4  & 5.36 & 91.4 & 3.21 & & & & 1.89 & 1.98 & \\
430   & 8.4  & 89.3 & 2.3  & 5.36 & 91.4 & 3.22 & & & & 1.94 & 1.98 & \\
435   & 5.8  & 91.5 & 2.8  & 5.36 & 91.4 & 3.22 & & & & 1.97 & 1.98 & \\
440   & 4.4  & 91.8 & 3.8  & 5.36 & 91.4 & 3.33 & & & & 1.99 & 1.98 & \\
445   & 3.9  & 90.0 & 6.1  & 5.36 & 83.6 & 11.2 & & & & 2.02 & 2.06 & \\
448   & 4.2  & 87.5 & 8.2  & 5.36 & 83.6 & 11.2 & & & & 2.04 & 2.06 & \\
450   & 4.0  & 88.3 & 7.7  & 5.36 & 75.5 & 19.3 & & & & 2.04 & 2.14 & \\
455   & 4.3  & 86.0 & 9.7  & 5.36 & 73.0 & 21.9 & & & & 2.05 & 2.17 & \\
460   & 4.3  & 84.6 & 11.1 & 5.36 & 71.6 & 23.3 & & & & 2.07 & 2.18 & \\
465   & 4.8  & 82.1 & 13.1 & 5.36 & 70.3 & 24.6 & & & & 2.08 & 2.19 & \\
471   & 4.3  & 83.7 & 12.0 & 5.36 & 69.8 & 25.1 & & & & 2.08 & 2.20 & \\

 \hline
\end{tabular}
\\
\caption{Our experimental and theoretical branching ratios for
  formation of N$^+$, N$^{2+}$, and N$^{3+}$ ions, following
  photoionization of atomic N near the 1{\it s} ionization
  threshold. The estimated combined statistical error and systematic
  error on the experimental results is 3$\%$, and approximately 50 meV
  for the photon energy, except for the starred data where the
  estimated error is 50$\%$. The theoretical values of \citet{Kaas93a}
  are included for comparison.}
\label{tab:ratios}
\end{sidewaystable}

\newpage

\end{document}